\documentstyle[aps,epsf,prc]{revtex}

\begin{document}
\title{Neutron electric form factor at large momentum transfer.}
\author{Egle Tomasi-Gustafsson}
\address{\it DAPNIA/SPhN, CEA/Saclay, 91191 Gif-sur-Yvette Cedex, 
France}

\author{Michail P. Rekalo \footnote{ Permanent address:
\it National Science Center KFTI, 310108 Kharkov, Ukraine}}
\address{Middle East Technical University, 
Physics Department, Ankara 06531, Turkey}

\maketitle

\def\gms{$G_{Ms}$}
\def\gmp{$G_{Mp}$}
\def\gmn{$G_{Mn}$}
\def\ges{$G_{Es}$}
\def\gep{$G_{Ep}$}
\def\gen{$G_{En}$}

\begin{abstract}
Based on the recent, high precision data for elastic electron scattering from 
protons and deuterons, at relatively large momentum transfer $Q^2$, 
we determine the neutron electric form factor up to $Q^2=3.5$ GeV$^2$. 
The values obtained from the data (in the framework of the nonrelativistic 
impulse 
approximation) are larger than commonly assumed and 
are in good agreement with the 
Gari-Kr\"umpelmann parametrization of the nucleon electromagnetic form factors.
\end{abstract}

\vspace*{.2true cm}

pacs numbers: {25.30.-c, 25.30.Bf, 21.45.+v}
\vspace*{.2true cm}

The internal structure of hadrons, their charge and magnetic distributions, can 
be 
conveniently described in terms of form factors.
Elastic electron-hadron scattering  is the traditional way  
to experimentally determine the electromagnetic form factors, and it allows a 
direct comparison with the theory. 

Direct measurements of the electric neutron form factor have been recently made 
possible, at transfer 
momenta $Q^2<$1 GeV$^2$, via inelastic electron scattering by deuteron or 
$^3\!He$. These 
experiments require not only a polarized beam but either a polarized target 
or the measurement of the polarization of the outgoing 
neutron (for a recent update see \cite{Wa99} and refs. herein).

Having high precision data on the differential cross section for $ed-$ elastic 
scattering, and assuming a 
reliable model for their description,  one can extract, in principle, the 
dependence of the electric neutron form factor \gen\  on the momentum transfer 
$Q^2$. Such a procedure  has been carried out in ref. \cite{Pl90}, up to 
$Q^2$=0.7 GeV$^2$. The purpose of this paper is to extend such an analysis 
at larger $Q^2$. This is motivated by the fact that 
the  elastic $ed$-scattering data extend up to $Q^2$=6 GeV$^2$, with high 
precision \cite{Al99} and the recent data on the proton electric form factor 
\cite{Jo00} (which have been obtain by the recoil proton polarization 
measurement following an idea suggested more than 30 
years ago \cite{Re68}) extend up to $Q^2$=3.5 
GeV$^2$.  A large program is under way at the Jefferson 
Laboratory (JLab) to improve the accuracy of the experimental data for the 
proton and  deuteron elastic electromagnetic form 
factors up to relatively large values of momentum 
transfer \cite{Al99,Jo00,Ab99}, to measure 
the neutron electric form factor \cite{Madey} and the electromagnetic transition 
form 
factors of the nucleon resonances 
\cite{Bu95}.

Common assumptions, in the previous calculations of deuteron electromagnetic 
form factors, 
are that the proton electric form factor, \gep , follows a 
dipole-like $Q^2$ dependence (resulting from an exponential charge 
distribution), 
and that \gen\ is negligible (in the 
region of space-like momentum transfer).

The large sensitivity to 
the nucleon form factors of the models which 
describe the light nuclei structure, particularly the deuteron, was carefully 
studied in \cite{Ar80}, and 
it was pointed out that the disagreement between the relativistic impulse 
approximation and the existing data [up to $Q^2$=4 GeV$^2$] could be 
significantly reduced if \gen\  were different from zero. 

On the other hand, recent studies focus primarily on other 
ingredients of the deuteron structure, such as the choice of the deuteron wave 
function or specific corrections like meson exchange currents (MEC), 
relativistic effects, and 
six-quark components in the deuteron (for a review see, for instance, 
\cite{Schiavilla}). 

Here we  take another approach: we look for a consistent explanation of 
JLab data about $ed-$ and $ep-$ 
scattering in the framework of the impulse approximation (IA) for 
$ed$-scattering, and  derive the 
neutron electric form factor. A discussion of the role of the corrections to IA, 
in particular the MEC, will follow.

In the framework of non relativistic IA,  where the calculation of the  deuteron 
electromagnetic form factors is straightforward, only two ingredients are 
required: the S- and 
D-components of the deuteron wave function and the electromagnetic form factors 
of the nucleons, which are considered as free ones, without off-shell mass 
effects.

Let us recall here some useful formulas (for a complete derivation, see 
\cite{Ja56}). 
The  differential cross section for $ed$ elastic scattering 
can be expressed in terms of two structure functions, $A(Q^2)$ and 
$B(Q^2)$, in one photon approximation\footnote{For a recent discussion of the validity of one-photon exchange in this 
momentum range, see ref. \cite{Re99}}:
$$\displaystyle\frac{d\sigma}{d\Omega}=
\left (\displaystyle\frac{d\sigma}{d\Omega}\right )_0 \cdot
{\cal S},~~{\cal S}= A(Q^2)+B(Q^2) ~\tan^2(\theta_e/2)
$$
with
$$
\left (\displaystyle\frac{d\sigma}{d\Omega}\right)_0=
\frac{\alpha^2~cos^2(\theta_e/2)E'}{4sin^4(\theta_e/2)E^3},
$$
where $E$ ($E'$)  is the electron beam (the scattered electron) energy. 
The structure functions $A$ and $B$ can be expressed in terms of the three 
deuteron form factors, $G_c$ 
(electric), $G_m$ (magnetic) and $G_q$ (electric quadrupole) as:
$$
A(Q^2) =G_c^2(Q^2)+ \frac{8}{9} \tau^2 G_q^2(Q^2)+\frac{2}{3} \tau 
G_m^2(Q^2),
$$
\begin{equation}
B(Q^2) = 
\frac{4}{3} (1+\tau) \tau G_m^2(Q^2),~\tau =\frac{Q^2}{4M^2_d}
\end{equation}
where $M_d$ is the deuteron mass.

In order to disentangle the three form factors it is necessary to measure the 
cross section  at least at two different angles for a fixed $Q^2$ (the 
Rosenbluth separation), and some polarization observables. In case of an 
unpolarized beam and 
target, the outgoing deuteron is tensor polarized and the components of the 
tensor polarization $t_{2,i}$ ($i=0-2$) give useful combinations of form 
factors. In particular $t_{20}$ is sensitive to $G_c$ and $G_q$:
\begin{equation}
t_{20}=-\displaystyle\frac{1}{\sqrt{2} {\cal S} }
\left \{ \displaystyle\frac{8}{3}\tau G_c G_q + 
\displaystyle\frac{8}{9} \tau^2 
G_q^2+  \frac{1}{3} \tau
\left [1+2 (1+\tau )\tan^2 (\theta_e/2 )\right ] G_m^2 \right \}
\end{equation}
In the non relativistic IA, the deuteron form factors
depend only on the deuteron wave function and on nucleon form factors:
$$
G_c=G_{Es}C_E,~~G_q=G_{Es}C_Q,
$$
\begin{equation}
G_m=\displaystyle\frac{M_d}{M_p}\left 
(G_{Ms}C_S+\displaystyle\frac{1}{2}G_{Es}C_L\right ),
\end{equation}
where $M_p$ is the proton mass, $G_{Es}=G_{Ep}+G_{En}$ and  
$G_{Ms}=G_{Mp}+G_{Mn}$ are the charge and 
magnetic isoscalar nucleon form factors, respectively. The terms 
$C_E$, $C_Q$, 
$C_S$, and $C_L$ describe the deuteron structure and can be calculated from the 
deuteron $S$ and $D$ 
wave functions, $u(r)$ and $w(r)$ \cite{Ja56} :
$$C_E={\int }_0^{\infty}dr~j_0\left( 
\frac{Qr}2\right) \left[ u^2\left( r\right) +w^2( r)
\right], $$
$$C_Q=\frac{3}{\sqrt{2}\eta}{\int }_0^{\infty}dr~j_2\left( 
\frac{Qr}2\right) \left[ u( r) -\frac{w( r)}{\sqrt{8}}\right] w(r),  $$
\begin{equation}
C_S={\int }_0^{\infty}dr \left[ u^2( r) -\frac{1}2w^2( r)
\right ]j_0\left( \frac{Qr}{2}\right) +
\frac{1}{2}\left [\sqrt{2}u( r)w(r)+w^2( r)\right ] j_2\left( 
\frac{Qr}{2} \right ),
\end{equation}
$$C_L=\frac{3}{2} \int_0^{\infty}dr~w^2( r)
\left [ j_0 \left ( \frac{Qr}{2} \right )+ j_2 \left ( \frac{Qr}{2}\right ) 
\right ]
$$
where 
$$j_0(x) =\frac{\sin x}{x},\ j_2( x) =
\sin x\left (\frac {3}{x^3}-\frac {1}{x}\right ) -3\frac{\cos x}{x^2}$$
are the spherical Bessel functions.
The normalization condition is $$
{\int }_0^{\infty}dr~\left[ u^2( r)+w^2( r)\right ]=1.$$

With the help of expressions (3) and (4), the formula for $A(Q)^2$, Eq. (1), 
can be inverted into a quadratic equation for \ges. Then \ges\  is calculated 
using the experimental values for $A(Q)^2$. We assume, for the magnetic nucleon 
form factors $G_{Mp}$ and $G_{Mn}$ 
the 
following dipole dependence, 
$$G_{Mp}(Q^2)/\mu_p=G_{Mn}(Q^2)/\mu_n=G_D,$$
with
$$\mu_p=2.79,~ 
\mu_n=-1.91,\mbox{~and~}
G_D =
\displaystyle\frac{1}
{\left [1+\displaystyle\frac{Q^2}{0.71 ~GeV^2 }\right ]^2},$$
which is in agreement with the existing data at a 3\%  level, up to
$Q^2\simeq$ 10 GeV$^2$.

In Fig. 1 we illustrate the behavior of the different nucleon 
electric form factors: \ges,\  \gep\  and \gen. The nucleon isoscalar 
electric form factor, 
derived from different sets of deuteron data, decreases when $Q^2$ increases. 
The solid line represents the 
Gari-Kr\"umpelmann parametrization \cite{G-K} for \ges . The dipole behavior, 
which is 
generally 
assumed for the proton electric form factor is shown as a dotted line. 
We have approximated the last \gep\  data by a function of the form:
\begin{equation}
G_{Ep}=\displaystyle\frac{G_D}{ 1+\displaystyle\frac {Q^2}{m_x^2}}
\end{equation} with 
$m_x^2$=5.88 GeV$^2$, (thin dashed-dotted line). The 
new \gep\  data, which decrease faster than the dipole function, are also well 
reproduced by the Gari-Kr\"umpelmann parametrization (thick dashed line). 

The  electric neutron form factor can be calculated 
from the isoscalar nucleon form factor, assuming for \gep\  a dipole behavior 
(solid stars) or Eq. (5) (open stars). The last option leads to values for 
\gen\  which are in  
very good agreement with the parametrization 
\cite{G-K}. These results shows that the neutron form factor
is not going to vanish identically at large momentum transfer, but becomes more 
sizeable 
than predicted by other parametrizations, often used in the calculations 
\cite{Pl90,Galster} (thin dashed line).
Starting from $Q^2\simeq 2$ GeV$^2$  the form factor \gen\  becomes even 
larger than \gep . Let us mention that a recent 'direct' measurement \cite{Ro99} 
at 
$Q^2=0.67$ GeV$^2$ finds \gen =$0.052\pm 0.011\pm 0.005$  in agreement with 
the present values.

In order to test the coherence of this description, we plot in Fig. 2 the 
prediction of the IA model, based on the Paris wave function and 
the parametrization \cite{G-K}, together with a sample of the existing data on 
the deuteron observables, $A(Q^2)$, $B(Q^2)$ and $t_{20}$.

The agreement is qualitatively good, even at large momentum transfer,
for all the observables. Replacing the common dipole approximation with Eq. (5), 
based on the new data about \gep,  induces little 
effect on the $Q^2$ - dependence of
$t_{20}$ and the structure function $B(Q^2)$, but the structure function 
$A(Q^2)$ gets much 
smaller, up to 40\% at $Q^2=4$ GeV$^2$. This explains the important role 
played 
by the neutron electric form factor, which is therefore larger than previously 
assumed. 
Replacing the Paris wave function with other N-N potentials gives qualitatively 
similar results.
Introducing different corrections to the IA as relativistic effects, MEC, 
isobar,
six-quark contributions etc.. brings to results which are largely model 
dependent \cite{etg}. 
Let us mention that the $\gamma^*\pi^{\pm}\rho^{\mp}$-contribution, which is a 
good approximation for the isoscalar 
transition $\gamma^*\rightarrow \pi^+\pi^-\pi^0$ 
($\gamma^*$ is a virtual photon), is typically considered as the main 
correction to IA, necessary, in particular, to improve the description of 
the SF $A(Q^2)$ \cite{Ar75}. However the relative role of MEC is strongly model 
dependent \cite{Bu92} as the coupling constants for meson-NN-vertexes 
are not well known and arbitrary form factors are often added \cite{Ad64,VO95}.

It should be pointed out that the $\gamma^*\pi\rho$ vertex is of 
magnetic nature and its contribution to $A(Q^2)$ has to be of the same order of 
magnitude as the relativistic corrections.
The general spin structure of the $\gamma^*\pi\rho-$vertex can be written as: 
$\epsilon_{\mu\nu\rho\sigma}e_{\mu}k_{\nu}U_{\rho}q_{\sigma},$ where $e$ and $k$ 
($U$ and $q$) are the 4-vector of the photon ($\rho$-meson) 
polarization and corresponding 4-momentum. The equivalent 3-dimensional 
expression for the spin structure is: $\vec e\times\vec k\cdot\vec U$ (where 
$\vec k$ is the photon 3-momentum), which allows the absorption of real (or  
virtual) $M1$ photon, only. A consequence is that the main contribution of 
$\gamma^*\pi\rho-$MEC concerns
$B(Q^2)$, which is proportional to the square of the deuteron magnetic form 
factor,  and not $A(Q^2)$, which mostly depends on the deuteron electric charge 
and quadrupole form factors.  

In conclusion, the  description of the deuteron electromagnetic structure in the 
framework of IA is now possible up to large momentum transfer ($Q^2\simeq$ 3.5 
GeV$^2$ ) and it  predicts a saturation of the isoscalar electric nucleon 
form factor by the neutron electric form factor at large $Q^2$. 
This result is consistent with the predictions of \cite{G-K}. There is no strong 
theoretical background 
in the ansatz \gen =0, often used in the literature concerning deuteron, in the 
considered region of space-like momentum transfer. The forthcoming data about 
\gen, planned at JLab  up to $Q^2$=2 GeV$^2$, \cite{Madey}
will be crucial in this respect. The large sensitivity of the 
deuteron structure to the nucleon form factors 
shows the necessity to reconsider the role of meson exchange currents
in the deuteron physics at large momentum transfer.

We  
thank  L. Cardman and  C.F. Perdrisat for interesting 
discussion on the last JLab results and on the recent \gep\  data, S. Platchkov,  
for his remarks on the extraction 
of neutron form factors and deuteron physics. We acknowledge the very 
positive and encouraging comments from S. Brodsky. Special thanks are due to M. 
Mac Cormick, for a careful reading of the manuscript.


\clearpage

\begin{figure}
\vspace*{-2truecm}
\begin{center}
\mbox{\epsfxsize=14.cm\leavevmode \epsffile{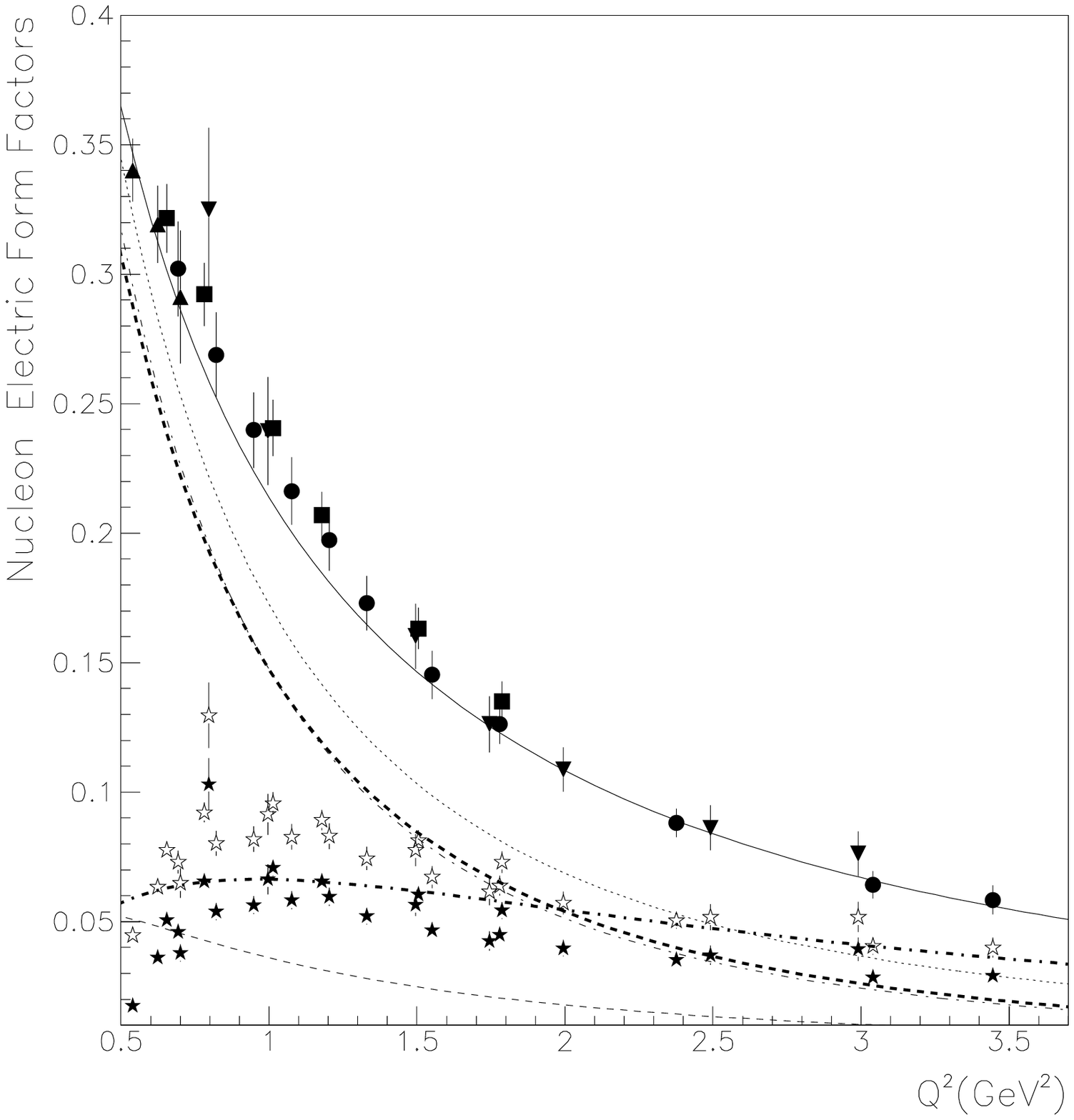}}
\end{center}
\vspace*{-2truecm}
\caption{Nucleon  electric form factors as functions of 
the momentum transfer $Q^2$. in the framework of IA with Paris potential.
Isoscalar electric  form factors are derived from  the deuteron elastic 
scattering data: \protect\cite{Pl90} (solid triangles), \protect\cite{Al99} 
(solid circles), \protect\cite{Ab99} (solid squares), and \protect\cite{Ar75} 
(solid reversed triangles). 
The  electric neutron form 
factors are shown as solid stars when calculated from the dipole
representation of \gep\  (dotted line) and open stars when 
Eq. (5) is taken for \gep\  (thin dashed-dotted line).  The parametrization  
\protect\cite{G-K} is shown for \ges (solid line), for \gen\  (thick 
dashed-dotted 
line) and for \gep\  (thick dashed line). The thin dashed line is the 
parametrization \protect\cite{Galster} for \gen.}
\end{figure}

\begin{figure}
\begin{center}
\mbox{\epsfxsize=14.cm\leavevmode \epsffile{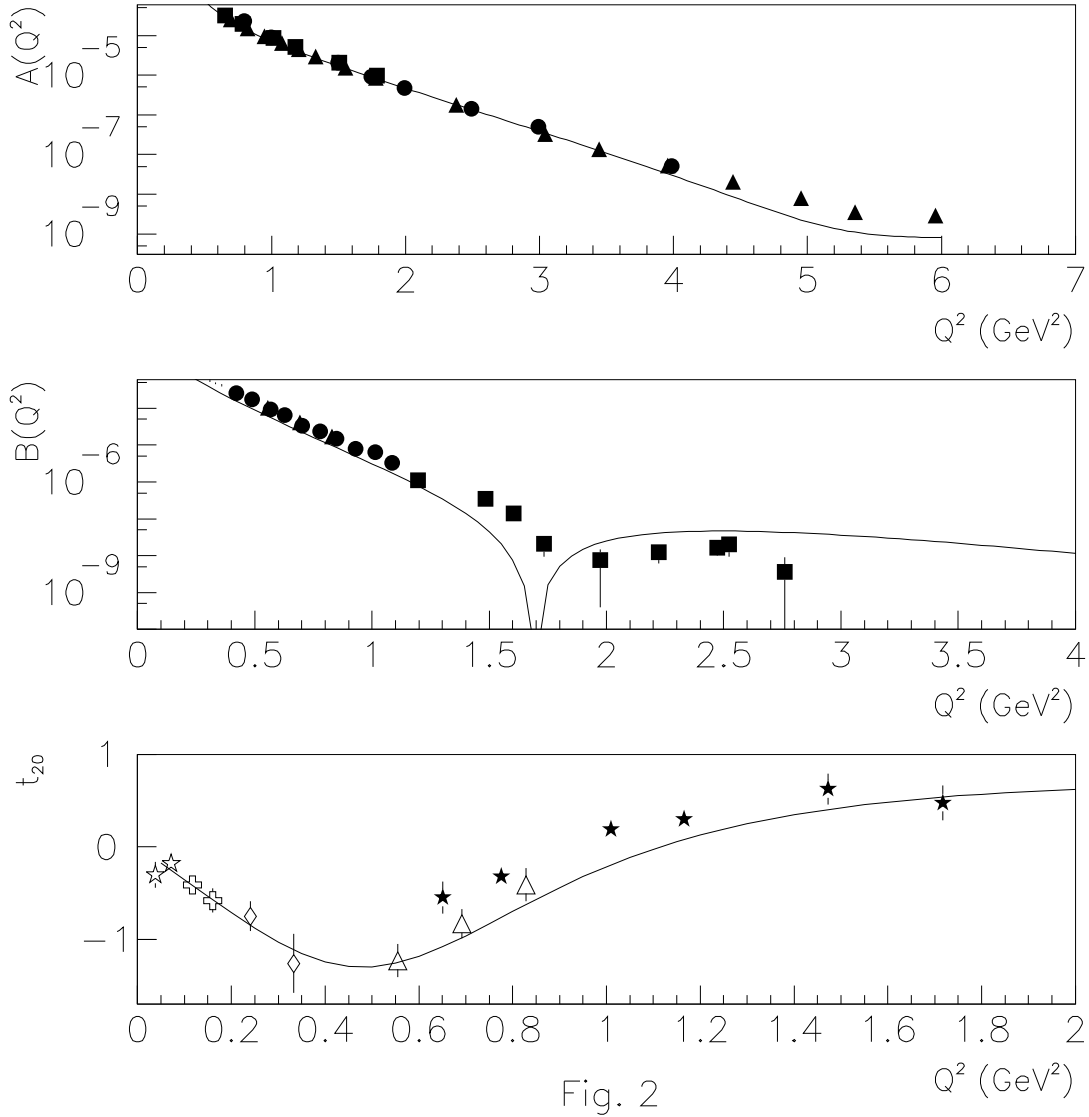}}
\end{center}
\caption{Impulse approximation prediction for the structure functions $A(Q^2)$, 
$B(Q^2)$, and  the 
tensor deuteron polarization $t_{20}$. The $A(Q^2)$ data are 
from 
\protect\cite{Al99,Ab99,Ar75}, the $B(Q^2)$ data are from \protect\cite{Bo90}, 
the $t_{20}$ data 
are from \protect\cite{t20}. }
\end{figure}

\end{document}